# Alex Müller, the High-$T_\mathrm{c}$ Field-Effect Transistor and Electric-Field Gated Quantum Materials

J. Mannhart

Max Planck Institute for Solid State Research

Heisenbergstr. 1, 70569 Stuttgart, Germany

*Abstract*

Alex Müller and Georg Bednorz are widely recognized for their trailblazing discovery of high-temperature superconductivity and their groundbreaking research on SrTiO$_3$. In comparison, their substantial contributions to inventing the high-$T_\mathrm{c}$ superconducting field-effect transistor remain relatively unknown. Nevertheless, their efforts were crucial in developing the electric field effect into a valuable tool for studying a broad spectrum of complex materials. This article provides a brief overview of these developments and of the current status in this field, with a particular focus on Alex Müller's visionary role in advancing the field following the discovery of high-temperature superconductivity.

*Introduction*

The field-effect transistor (FET), invented by Julius Edgar Lilienfeld in 1925, is ubiquitous in modern digital and analog electronics. Semiconducting FETs use an electric field generated by a gate electrode G to alter the carrier density in a semiconducting DS channel that connects the drain D and source S electrodes, thereby controlling the channel's conductance. In the history of humanity, no product has been produced in greater numbers. Intel alone fabricates more than $10^{18}$ silicon-based FETs annually, and that number continues to grow.

Today, electric fields are used not only to modify the conductivity of standard semiconductors such as silicon, they also serve as a valuable tool of fundamental research to investigate and alter the properties of complex materials. In some materials, applied electric fields can even induce novel behavior that is distinct from the original material properties. Numerous samples of a wide range of materials have been examined, including complex oxides, materials with electronic correlations, 2D materials, and organic systems. Electric fields are generated in such studies by one of several methods. Standard gate stacks comprising either dielectric or ferroelectric gate insulators and gate electrodes are commonly used to generate the gate fields, as are liquid or solid electrolytes, or surface charges supplied either by thin films or by ions on the sample surface. The discovered and explored phenomena encompass field-induced superconductivity, magnetism, and microstructural switching.

This volume is dedicated to honoring Alex Müller's scientific achievements. As such, it would not be complete without highlighting his unfortunately lesser-known role in promoting the development of superconducting FETs that utilize high-$T_c$ cuprates as DS channels, as well as his dedication in supporting their experimental realization and exploration. For decades, these far-reaching efforts involved also numerous further scientists, all of whom I cannot mention by name within the scope of this article, although they certainly also deserve enormous credit for the success of the endeavors described below. This paper is intended to especially highlight Alex Müller's instrumental contributions as instigator of these research efforts.

*Studies of the Electric-Field Effect in Superconductors Prior to the Discovery of High-$T_c$ Cuprates*

Long before the discovery of high-$T_c$ cuprates, researchers explored electric fields as a tool to modify the properties of DS channels. The first recorded experiments of this kind were performed by Emil Bose in 1906, who applied gate fields to macroscopic platinum strips ([1], see also [2]). Unfortunately, these efforts did not yield significant results because the carrier density of conventional metals is of the order of $10^{23}$ cm$^{-3}$. Their Thomas–Fermi screening length is therefore so small that mobile charges shield the electric fields directly at the channel's surface. This prevents the field from penetrating the channel and altering its bulk properties. With this understanding, studies proceeded to investigate the effects of applied electric fields on properties of thin films or their surfaces.

Glover and Sherill applied the electric-field effect to change the critical temperature $T_c$ of superconducting thin films as early as 1960 [3]. They measured a shift of $T_c$ of $\approx 100$ μK for In and Sn films. Intrigued by the concept of controlling superconductivity through external electric fields, researchers then conducted numerous experiments to enhance these effects. In the 1980's, oxide and chalcogenide superconductors such as doped $SrTiO_3$ [4], $Ge_{1-x}Te$ [4], InO [5] and $Ba(Pb, Bi)O_3$ [6] were appreciated to have low carrier densities and to therefore yield relatively large electric-field effects on the superconducting behavior. To achieve superconductivity with small carrier densities, superconducting FETs that used proximitized semiconductors such as InAs as DS channel materials were furthermore explored [7].

Although only few researchers conducted these studies from 1906–1986, many of their groundbreaking ideas presaged future concepts. However, owing to the limitations of the materials available at that time, the observed effects were minor. Electric-field-effect studies were a niche topic of research and went largely unnoticed by the broader scientific community. Early summaries of superconducting field-effect devices studied before the discovery of high-temperature superconductivity (HTS) can be found in [8] and [9].

*High-$T_c$ Field-Effect Transistors*

The discovery of HTS in cuprates by Georg Bednorz and Alex Müller in 1986 [10] revealed an intriguing new family of candidate materials for channels in superconducting FETs. The carrier density of superconducting cuprates is in the range of $10^{21}$–$10^{22}$ cm$^{-3}$, which is significantly lower than that of standard metallic superconductors, yet higher than that of typical doped semiconductors.

Familiar with work on GaAs-based modulation-doped FETs (MODFET) conducted at the IBM Research Laboratory in Rüschlikon, Alex Müller together with Peter Wolf, a member of the IBM MODFET team, immediately envisioned the potential of high-$T_c$ cuprates for FET applications. In the fall of 1987, they pointed out that high-$T_c$ cuprates have five advantages for FETs: a relatively low carrier density, a very small out-of-plane coherence length, a high critical current density $J_c$ of epitaxial films, a short switching time and, of course high-$T_c$. Together with Praveen Chaudhari of the IBM Research Center in Yorktown Heights, Müller and Wolf filed a European Patent Application entitled "*Field Effect Device with a Superconducting Channel*" in January 1988, see Fig. 1 [11]. This application documents that those three researchers were thinking beyond individual field-effect transistors: They envisioned entire electronic chips made of cuprate superconductors. They recognized that the high $J_c$ of epitaxial $YBa_2Cu_3O_{7-x}$ films made it possible to use such films not only as DS channels but also as superconducting interconnects. A chip could be fabricated primarily from cuprates or a combination of cuprates and semiconductors, with the latter's performance enhanced by the low operating temperature required to match the cuprate $T_c$.

Bringing this vision to fruition became one of Alex Müller's favorite research interests following the HTS discovery. The prevailing sense of optimism at that time fostered visionary concepts and insights, as evidenced by the material parameters specified for such devices. The DS channel was expected to consist of a single monolayer of a high-$T_c$ cuprate such as $YBa_2Cu_3O_{7-x}$ with a carrier density of $\approx 10^{21}$ cm$^{-3}$ and a $J_c$ value greater than $10^7$ A cm$^{-2}$. A 5-nm-thick film of $SrTiO_3$ was to be epitaxially grown as a gate insulator. In such a configuration (Fig. 2), a gate voltage of only a few volts was predicted to drive the $YBa_2Cu_3O_{7-x}$ channel across its entire thickness into complete depletion, thus changing the channel from superconducting to fully insulating. Today, we know it is unfortunately impossible to fabricate heterostructures that fulfill these specifications.

In fact, in 1988, thin-film heterostructures comprising high-$T_c$ cuprates had not yet been fabricated at all. Alex Müller's vision to utilize such heterostructures therefore provided one of his reasons to push for the installation of an oxide MBE system at IBM in Rüschlikon, envisaging this system to quickly realize HTS FETs. However, even today, establishing the growth of heterostructures of the required complexity by MBE would demand a substantial time frame, let alone in 1989. Therefore, the MBE FET project was rather slow in taking off. Coincidentally, unaware of IBM's interests in superconducting FETs, but thinking along similar lines, I applied that year for a position at IBM Rüschlikon with a proposal to fabricate *in situ* HTS-insulator-normal metal devices with sputtering and pulsed laser deposition. Such devices promised many opportunities for research, including their use as tunnel junctions and field-effect devices. I was hired by Alex Müller and Georg Bednorz, and supported to set up such a cluster system. After a few months, we were joined by Darrell Schlom, who was succeeded by Jörg Ströbel followed by Toni Frey, then Bernd Mayer, and finally Hans Hilgenkamp as postdocs.

We explored the phase space of possible device configurations. Alex Müller was delighted when our studies revealed that $SrTiO_3$, preferably 15% Ba-doped, was indeed the most suitable gate insulator due to its high dielectric constant. To avoid any degradation of the ultrathin high-$T_c$ films by growing $SrTiO_3$ films on their surface, we developed back-gate configurations that utilized conducting Nb-doped $SrTiO_3$ substrates as gates. These crystals were grown by Georg Bednorz in his home-built zone-melting furnace. Because we were searching for metallic substrate materials for the use as back-gates, he also suggested that for this purpose we explore $Sr_2RuO_4$ single crystals. Frank Lichtenberg then grew these crystals by zone melting. We succeeded in epitaxially growing good $YBa_2Cu_3O_{7-x}$ films on those crystals [12], but found that $SrTiO_3$ gate insulators were always conducting, presumably due to in-diffusion of Ru. Amazingly, when Yoshi Maeno cooled Frank's crystals to dilution refrigerator temperatures, he discovered they were superconducting [13]. As expected, our FET studies of the Nb-doped $SrTiO_3$ back-gates revealed that we could indeed shift the $T_c$ of $YBa_2Cu_3O_{7-x}$ or $Bi_2Sr_2Ca_1Cu_2O_{8+x}$, albeit initially by only $\approx 100$ mK [14]. Nevertheless, the effects were reproducible and were found not to be caused by another mechanism such as electrostriction in the SrTiO$_3$ substrates or by oxygen motion, but at least to a large degree by the electric field [14, 15]. Thus, pursuing Alex Müller's proposal, we succeeded in fabricating the first HTS FET devices [14]. Of course, these were far from being useful for practical applications.

By working through a series of design and fabrication improvements, during which Alex Müller contributed to three additional patent families, we increased the $T_c$ shifts to ≈3 K. Unfortunately, we did not manage to achieve greater shifts because it was simply impossible with the means available at that time to grow an ultrathin YBa$_2$Cu$_3$O$_{7-x}$ or Bi$_2$Sr$_2$Ca$_1$Cu$_2$O$_{8+x}$ layer encapsulated in a FET with useful values of $T_c$ and $J_c$. Moreover, the voltage amplification of these devices was reduced by the electric-field dependence of SrTiO$_3$ permittivity, as explored by Hans-Martin Christen [16].

Even so, we did succeed in fabricating Josephson-junction FETs by incorporating grain boundaries into the channels, thereby obtaining $T_c$ shifts up to 25–30 K [17, 18]. However, grain boundaries limited the potential of large $T_c$ shifts by preventing the channels from achieving a $J_c$ value of $10^7$ A cm$^{-2}$.

It is noteworthy that, although the microscopic mechanism behind changing the $T_c$ of a superconductor by an applied electric field was well understood by Alex Müller and other insiders, this understanding was not yet widespread in the broader superconducting community. Consequently, our work was dismissed by some, based on the misconception that, owing to its infinite conductivity, a superconductor would instantaneously shield itself perfectly from an applied electric field!

These studies yielded two main results. First, they demonstrated that Alex Müller's concept of FETs with high-$T_c$ cuprate superconducting DS channels was fundamentally viable, but that the specific material parameters of the cuprates prevented the envisioned potential for digital electronics to be achieved.

Second, they demonstrated the potential of gated phase transitions for basic studies of condensed-matter physics and device applications. As a result, numerous other research groups quickly focused on this emerging topic (for an overview see, *e.g.*, [19]), spearheaded by Xiaoxing Xi, Venky Venkatesan and colleagues at the University of Maryland [20]. The group of Art Hebard and Anthony Fiory had indeed continued their field-effect work conducted in pre-HTS times and had begun early on to explore the effect of electric fields on YBa$_2$Cu$_3$O$_{7-x}$ surfaces [21]. A few years later, apparently superb results with superconducting FETs were reported by J.H. Schön, which appeared to outclass our work. This must have been especially frustrating for Alex Müller.

*Electric-Field-Effect Devices as Tools for Fundamental Research*

Within a few years, applied electric fields became widely appreciated as valuable tools to tune and switch the properties of quantum materials. Field-effect studies, once considered niche experiments, quickly became a topic of mainstream research. Thanks to the numerous outstanding activities that emerged in this process, experimental approaches expanded. For example, ionic gating allowed for greater polarizations than could be achieved with dielectric or ferroelectric gate insulators [22]. Additionally, nanoscale local gating on lithographically defined structures or via scanning probes proved successful for mesoscopic experiments involving complex materials and interfaces (see, *e.g.*, [23]).

Gate fields have been extensively employed to modify a wide range of materials beyond HTS cuprates (see, *e.g.*, [19], [24]), which may be broadly categorized today as "quantum materials". The gating of compounds with strong electron correlations is a heavily researched area, along with investigations of field-effect gating of interface electron systems [25] and the use of gate fields to induce superconductivity in inherently non-superconducting materials. This remarkable effect was first achieved in $KTaO_3$ by Ueno and coworkers [26]. Recently, electric fields applied to ferroelectric $MoTe_2$ bilayers have been used to switch between the nonsuperconducting and superconducting states of that compound [27]. Electric-field-induced metal–insulator phase transitions (*e.g.*, in $VO_2$ [28]) and electric-field-controlled quantum-phase transitions [29] are further examples of the growing field of electric-field tuning of quantum materials.

*Electric-Field-Effect Devices That Do Not Use Standard Semiconductors as DS Channels*

Research activities that employ the electric-field effect to tune the properties of quantum materials are now extensive in the materials space and comprise numerous types of investigations. It is to be expected that several of these studies will lead to important device applications.

In my opinion, the most intriguing device category are FETs that utilize gated phase transitions, as envisioned by Alex Müller for HTS FETs. In phase-transition FETs, some being called Mott transistors, an electric field induces a phase transition, examples of which were given above. If a phase transition is triggered, an ensemble of the electrons comprising numerous electron charges switches as a whole in a nonlinear process, hereby responding with high sensitivity to the applied gate field. Owing to the high effective charge involved in the switching process, these FETs have the potential to feature a small sub-threshold slope. Additionally, the switching of a function, such as zero resistance for the HTS FET, holds great promise for applications. It is also possible to induce such a switching by means other than electric fields, such as current injection explored for memristor applications. Brief overviews of phase-transition FETs are presented in [19,30].

*Summary*

It is undeniable that the research area of electric-field-gated complex oxides would have burgeoned at some point, even without Alex Müller's visionary concepts. However, as history unfolded, the ef-forts to achieve HTS FETs, envisaged and co-initiated by Alex Müller, played a key role in inspiring and encouraging the development of this field.

Although cuprate FET technology has not evolved into commercial products, it has significantly enhanced our understanding of quantum-matter heterostructures as well as our proficiency in their epitaxial growth [31]. This outcome has launched a thrilling, expanding, and prolific field of fundamental research: the investigation of quantum materials through electric-field gating.

As Praveen Chaudhari once quipped, "With his well-developed nose, Alex smells promising new areas in science". This characterization definitely proved correct for the electric-field-induced switching of HTS superconductivity.

Nature 456 (2008) 624-627,
https://doi.org/10.1038/nature07576.

30. Z. Yang, C. Ko, S. Ramanathan,
    Oxide electronics utilizing ultrafast metal-insulator transitions,
    Annu. Rev. Mater. Res. 41 (2011) 337-367,
    https://doi.org/10.1146/annurev-matsci-062910-100347.

31. H. Boschker, J. Mannhart,
    Quantum-matter heterostructures,
    Annu. Rev. Condens. Matter Phys. 8 (2017) 145-164,
    https://doi.org/10.1146/annurev-conmatphys-031016-025404.

**Figure Legends**

*Figure 1*

First page of the European Patent Publication on the HTS FET [11], filed on Jan. 15, 1988 by P. Chaudhari, C.A. Müller, and H.P. Wolf.

*Figure 2*

Operation principle of an HTS FETs. The sketches show cross-sectional cuts through a top-gate device in the "off" (left) and "on" (right) states, see also at the bottom of Fig. 1. This image was taken from the patent application shown in Fig. 1 (11). The numbers 22, 23, 25 mark the S, D, and G electrodes, respectively, the gate insulator is labelled as 24, the DS channel as 21.

(54) **A field-effect device with a superconducting channel.**

(57) A field-effect structure, formed on a substrate (20) and comprising a channel (21) with source (22) and drain (23) as well as a gate (25) that is separated from the channel by an insulating layer (24). The channel is made of a high-$T_c$ metal-oxide superconductor, e.g., YBaCuO, having a carrier density of about $10^{21}/cm^3$ and a correlation length of about .2 nm. The channel thickness is in the order of 1 nm, it is single crystalline and oriented such that the superconducting behaviour is strongest in the plane parallel to the substrate. With a signal of a few Volt applied to the gate, the entire channel cross-section is depleted of charge carriers whereby the channel resistance can be switched between "zero" (undepleted, superconducting) and "very high" (depleted).

FIG. 3A
OFF

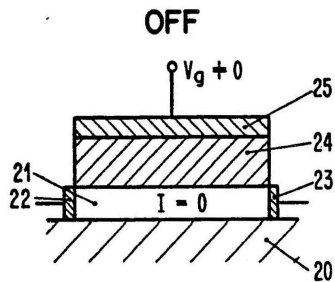

FIG. 3B
ON

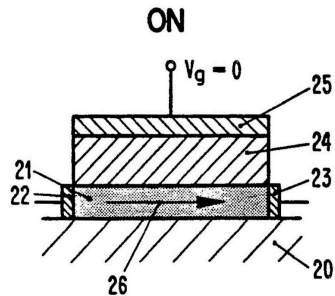

Figure 1

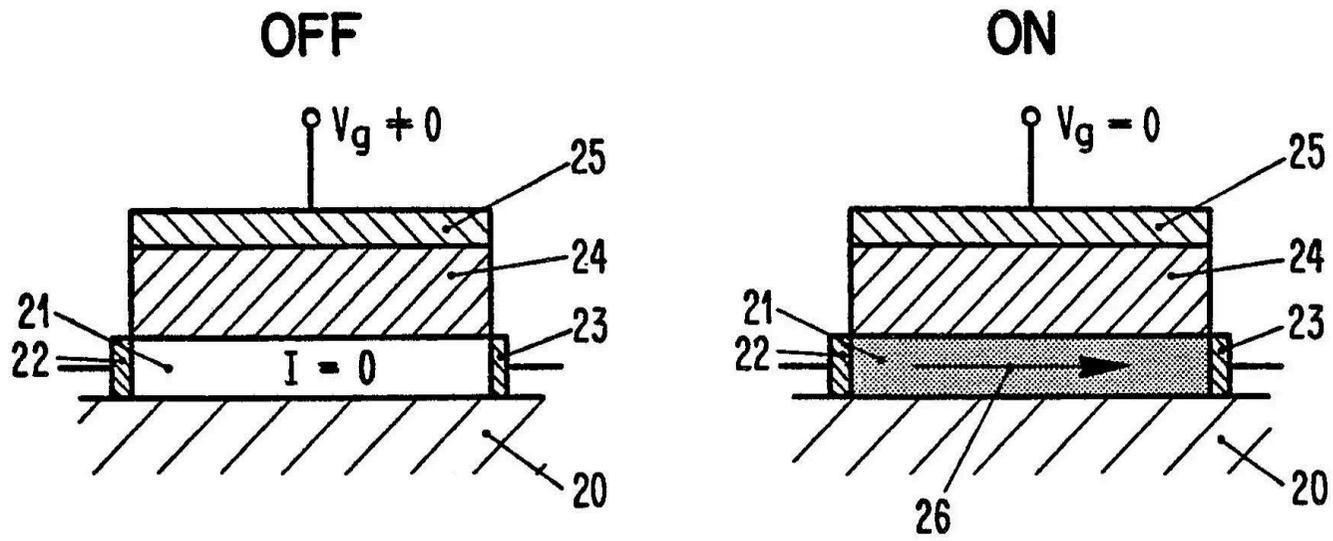

Figure 2